\documentclass{article}
\usepackage{arxiv}

\AtBeginDocument{%
  \providecommand\BibTeX{{%
    \normalfont B\kern-0.5em{\scshape i\kern-0.25em b}\kern-0.8em\TeX}}}

\usepackage[english]{babel}
\defineshorthand{"=}{\babelhyphen{nobreak}}
\useshorthands{"}
\usepackage{amsmath,amssymb,amsfonts}
\usepackage{algorithmic}
\usepackage{graphicx}
\usepackage{textcomp}
\usepackage{xcolor}
\usepackage[utf8]{inputenc}
\usepackage{array}
\PassOptionsToPackage{hyphens}{url}\usepackage{hyperref}

\usepackage{flushend} 
\newif\iflastpage
\usepackage{etoolbox} 
\makeatletter 
\preto\clearpage{%
  \begingroup 
    \let\@outputdblcol\last@outputdblcol 
} 
\appto\clearpage{%
  \endgroup 
} 
\makeatother 

\usepackage{natbib}
\usepackage{tikz}
\usepackage{calc}
\usetikzlibrary{arrows,automata,matrix}
\usetikzlibrary{arrows,arrows.meta}
\usetikzlibrary{positioning,arrows}
\usetikzlibrary{shapes,arrows,fit,calc,positioning,automata}
\usepackage{float}
\usetikzlibrary{positioning}
\usetikzlibrary{arrows,arrows.meta}
\usetikzlibrary{decorations.pathreplacing}
\usetikzlibrary{calc}
\usetikzlibrary{arrows,positioning,automata,shadows,fit,shapes}
\usetikzlibrary{trees}
\usetikzlibrary{patterns}
\usepackage{enumitem}
\usepackage{xcolor}

\usepackage{amssymb}
\usepackage{pifont}
\newcommand{\cmark}{\ding{51}}%
\newcommand{\xmark}{\ding{55}}%
\newcommand*\circled[1]{\tikz[baseline=(char.base)]{%
            \node[shape=circle,fill=blue!20!white,inner sep=0.1pt] (char) {\footnotesize \textcolor{black}{#1}};}}
\newcommand*\circledg[1]{\tikz[baseline=(char.base)]{%
            \node[shape=circle,fill=black!40!white,inner sep=0.1pt] (char) {\footnotesize \textcolor{white}{#1}};}}
\newcommand*\circledp[1]{\tikz[baseline=(char.base)]{%
            \node[shape=circle,fill=yellow!20!white, inner sep=0.1pt] (char) {\footnotesize \textcolor{black}{#1}};}}
                     
\usepackage{textcomp}
\usepackage{xspace}
\let\OldTexttrademark\texttrademark
\renewcommand{\texttrademark}{\OldTexttrademark\xspace}%

\usepackage{color}
\usepackage{listings,chngcntr}
\definecolor{codegreen}{rgb}{0,0.6,0}
\definecolor{codegray}{rgb}{0.5,0.5,0.5}
\definecolor{codepurple}{rgb}{0.58,0,0.82}
\definecolor{backcolour}{rgb}{0.95,0.95,0.92}

\usepackage{multirow}
\newcommand{\cell}[2][c]{%
  \begin{tabular}[#1]{@{}c@{}}#2\end{tabular}}
  
\usepackage[nolist,nohyperlinks]{acronym}
\usepackage{cleveref}

\makeatletter
\newcommand*{\org@overidelabel}{}
\let\org@overridelabel\@verridelabel
\@ifpackagelater{acronym}{2015/03/21}{
  \renewcommand*{\@verridelabel}[1]{%
    \@bsphack
    \protected@write\@auxout{}{\string\AC@undonewlabel{#1@cref}}%
    \org@overridelabel{#1}%
    \@esphack
  }%
}{
  \renewcommand*{\@verridelabel}[1]{%
    \@bsphack
    \protected@write\@auxout{}{\string\undonewlabel{#1@cref}}%
    \org@overridelabel{#1}%
    \@esphack
  }%
}
\makeatother
\Crefformat{figure}{#2Figure~#1#3}
\Crefmultiformat{figure}{Figures~#2#1#3}{ and~#2#1#3}{, #2#1#3}{ and~#2#1#3}
\Crefformat{table}{#2Table~#1#3}
\Crefmultiformat{table}{Tabels~#2#1#3}{ and~#2#1#3}{, #2#1#3}{ and~#2#1#3}
\Crefformat{equation}{#2Equation~#1#3}
\Crefmultiformat{equation}{Equations~#2#1#3}{ and~#2#1#3}{, #2#1#3}{ and~#2#1#3}
\crefname{lstlisting}{listing}{listings}
\Crefname{lstlisting}{Listing}{Listings}

\usepackage[nolist,nohyperlinks]{acronym}
\begin{acronym}
    \acro{APT}{Advanced Persistent Threat}
    \acro{ARP}{Address Resolution Protocol}
    \acro{BMBF}{Federal Ministry of Education and Research}
    \acro{CIA}{Confidentiality, Integrity and Availability}
    \acro{CPU}{Central Processing Unit}
    \acro{DHCP}{Dynamic Host Configuration Protocol}
    \acro{DoS}{Denial of Service}
    \acro{DMA}{Direct Memory Access}
    \acro{ERP}{Enterprise Resource Planning}
    \acro{HAL}{Hardware Abstraction Layer}
    \acro{HMI}{Human Machine Interface}
    \acro{I2C}{Inter"=Integrated Circuit}
    \acro{ICMP}{Internet Control Message Protocol}
    \acro{ICS}{Industrial Control System}
    \acrodefplural{ICS}[ICSs]{Industrial Control Systems}
    \acro{IDE}{Integrated Development Environment}
    \acro{IDS}{Intrusion Detection System}
    \acro{IIoT}{Industrial Internet of Things}
    \acro{IO}{input/output}
    \acro{IoT}{Internet of Things}
    \acro{IP}{Internet Protocol}
    \acro{LED}{Light"=Emitting Diode }
    \acro{LwIP}{Lightweight IP}
    \acro{MAC}{Media Access Control}    
    \acro{MCU}{Microcontroller Unit}
    \acro{MES}{Manufacturing Execution System}
    \acro{MitM}{Man in the Middle}
    \acro{NSE}{Nmap Scripting Engine}
    \acro{OLED}{Organic Light Emitting Diode}
    \acro{PCB}{Printed Circuit Board}
    \acro{PLC}{Programmable Logic Controller}
    \acrodefplural{PLC}{Programmable Logic Controllers}
    \acro{PoC}{Proof of Concept}
    \acro{RAM}{Random"=Access Memory}
    \acro{ROM}{Read"=only Memory}
    \acro{RTOS}{Real"=time Operating System}
    \acro{SCADA}{Supervisory Control and Data Acquisition }
    \acro{SPI}{Serial Peripheral Interface}
    \acro{ST}{Structured Text}
    \acro{STM}{STMicroelectronics}
    \acro{SWD}{Serial Wire Debug}
    \acro{TCP}{Transmisson Control Protocoll}
    \acro{USB}{Universal Serial Bus}
\end{acronym}

\usepackage{filecontents}

\usepackage{graphicx}

\newcommand\copyrighttext{%
  \footnotesize \textcopyright Efficient Passive ICS Device Discovery and Identification by MAC Address Correlation -- First published in the Electronic Workshops in Computing series at
  6th International Symposium for ICS \& SCADA Cyber Security Research 2019 (ICS-CSR 2019)
  DOI: \href{http://dx.doi.org/10.14236/ewic/ICS2019.1}{10.14236/ewic/icscsr19.1}
}
\newcommand\copyrightnotice{%
\begin{tikzpicture}[remember picture,overlay]
\node[anchor=south,yshift=10pt] at (current page.south) {\fbox{\parbox{\dimexpr\textwidth-\fboxsep-\fboxrule\relax}{\copyrighttext}}};
\end{tikzpicture}%
}

\title{LICSTER -- A Low-cost ICS Security Testbed for Education and Research}

\author{
 Felix Sauer \\
 felix.sauer@hs-augsburg.de \\
 Hochschule Augsburg
 \And
 Matthias Niedermaier \\
 matthias.niedermaier@hs-augsburg.de \\
 Hochschule Augsburg
 \And
 Susanne Kießling \\
 susanne.kiessling@hs-augsburg.de \\
 Hochschule Augsburg
 \And
 Dominik Merli \\
 dominik.merli@hs-augsburg.de \\
 Hochschule Augsburg
}

\begin{document}

\maketitle

\begin{abstract}
    Unnoticed by most people, \acp{ICS} control entire productions and critical infrastructures such as water
    distribution, smart grid and automotive manufacturing.
    Due to the ongoing digitalization, these systems are becoming more and more connected in order to enable remote
    control and monitoring.
    However, this shift bears significant risks, namely a larger attack surface, which can be exploited by
    attackers.
    In order to make these systems more secure, it takes research, which is, however, difficult to conduct on productive
    systems, since these often have to operate twenty"=four"=seven.
    Testbeds are mostly very expensive or based on simulation with no real"=world physical process. 
    In this paper, we introduce LICSTER, an open"=source low-cost \ac{ICS} testbed, which enables researchers and students to get 
    hands"=on experience with industrial security for about 500~Euro.
    We provide all necessary material to quickly start \ac{ICS} hacking, with the focus on low"=cost and open"=source
    for education and research.
\end{abstract}

\copyrightnotice

\acresetall
\section{Introduction}
\label{sec:introduction}
New concepts brought by industry~4.0 start to drastically change the requirements for industrial plants.
Predictive maintenance for example, a concept where machines get serviced before they fail based on collected usage data,
demand for a high amount of data sources and pervasive communications.
\ac{ICS} devices such as \acp{HMI} or \ac{SCADA} systems gather the data within these environments, e.g. from sensors and actors.
Nowadays, most \ac{ICS} devices communicate via TCP/IP protocols with each other.
The rise in connected devices inherently increases the attack surface of a network and thus the systems within it.
To keep the evermore complex \acp{ICS} secure, it takes trained and well educated personnel as well as substantial research in this area.
However, it is risky to conduct research and training on productive plants, as minor disturbances can quickly lead to
costly incidents.
Due to this reason, testbeds for research and education are essential.

In principle, there are three types of testbeds: Virtualized, real"=world and hybrid approaches.
Just as there are different types, there are different tasks for which a testbed can be used.
For security scenarios and attacks on an \ac{ICS} in particular, a real"=world testbed incorporating a physical process
is preferred to wholly understanding the effects and attack vectors in a production environment.
Unfortunately, purchasing real industrial hardware for a testbed is very expensive and
particularly for education and research often not affordable.
Additionally, the proprietary devices prevent pervasive changes,
which makes research partly difficult.

In this paper, we present LICSTER, an open"=source, low"=cost \ac{ICS} testbed with the following contributions:
\begin{itemize}
    \item Testbed components for \textbf{about 500 Euro}, which is affordable by most researchers and students.
    \item A real"=world physical process controlled by an \ac{ICS},
          which enables to demonstrate and analyze the impacts of cyber attacks in the real"=world.
    \item The \textbf{feasibility} of the testbed is shown and ideas for \textbf{research} are discussed.
    \item The components are \textbf{open"=source} and \textbf{open"=hardware}, as far as possible. This allows a wide range of further research.
    \item We provide \textbf{attacker models and attacks} to understand threat scenarios in industrial environments.
\end{itemize}

The paper is structured as follows:
In \Cref{sec:concept}, the concept of LICSTER is presented.
\Cref{sec:implementation} describes the proposed implementation of the components.
The paper continuous with an evaluation in \Cref{sec:evaluation}, with a discussion about further training and research questions.
Eventually, \Cref{sec:conclusion} concludes this work.

\section{ICS Testbed}
\label{sec:concept}
Setting up a testbed is mostly not the final goal, but simply a tool to achieve a bigger objective.
That makes it crucial to have a clear understanding of that objective and its constraints before beginning to design a testbed.
Especially for an \ac{ICS} security testbed for education and research,
having clear attacker models and a prepared list of attack scenarios is valuable.

\subsection{\ac{ICS} Basics}
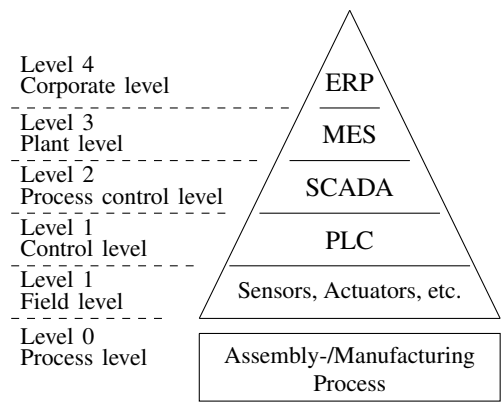
\begin{figure}[htb]
  \centering
\begin{tikzpicture}[node distance=1cm,
    auto
  ]
  
  \node[align=center, anchor=south] at (4,5.8) {ERP};
  \node[align=center, anchor=south] at (4,5.1) {MES};
  \node[align=center, anchor=south] at (4,4.4) {SCADA};
  \node[align=center, anchor=south] at (4,3.7) {PLC};
  \node[align=center, anchor=south] at (4,3) {\small Sensors, Actuators, etc.};
  \node[align=center, anchor=south] at (4,1.8) {\small Assembly-/Manufacturing \\ \small Process};
  \draw [-]  (2.0,2.9) -- (6.0,2.9);
  \draw [-]  (2.0,2.9) -- (4,7);
  \draw [-]  (6.0,2.9) -- (4,7);
  \draw [-]  (2.4,3.6) -- (5.6,3.6);
  \draw [-]  (2.8,4.3) -- (5.2,4.3);
  \draw [-]  (3.2,5.0) -- (4.8,5.0);
  \draw [-]  (3.6,5.7) -- (4.4,5.7);
  
  \draw [-]  (2.0,2.7) -- (6.0,2.7) -- (6.0,1.8) -- (2.0,1.8) -- (2.0,2.7);
  
  \draw [-, dashed] (-0.5,3.6) node[anchor=south west, align=left, text width=2.0cm]{\footnotesize Level 1 \\[-0.1cm] Control level}--(2.0,3.6);
  \draw [-, dashed] (-0.5,4.3) node[anchor=south west, align=left, text width=3.5cm]{\footnotesize Level 2 \\[-0.1cm] Process control level}--(2.4,4.3);
  \draw [-, dashed] (-0.5,5.0) node[anchor=south west, align=left, text width=2.0cm]{\footnotesize Level 3 \\[-0.1cm] Plant level} -- (2.8,5.0);
  \draw [-, dashed] (-0.5,5.7) node[anchor=south west, align=left, text width=2.5cm]{\footnotesize Level 4 \\[-0.1cm] Corporate level}-- (3.2,5.7);
  \draw [-, dashed] (-0.5,2.9) node[anchor=south west, align=left, text width=1.5cm]{\footnotesize Level 1 \\[-0.1cm] Field level} node[anchor=north west, align=left, text width=2.5cm]{\footnotesize Level 0 \\[-0.1cm] Process level} -- (1.6,2.9);
    
\end{tikzpicture}
    \caption{IEC 62264 Industrial Automation Pyramid}
\label{fig:lowcosticsIEC}
\end{figure}

The \cite{international2003iec} introduced a standard structure of \acp{ICS} in the IEC 62264, as illustrated in \Cref{fig:lowcosticsIEC}.

\textbf{\ac{ERP}} stands for the entrepreneurial task of planning and controlling resources
such as capital, personnel, operating resources and materials in a timely and appropriate manner.
A \textbf{\ac{MES}} refers to a process"=related level of a multi"=layered manufacturing management system.
\textbf{\ac{SCADA}} defines a system for monitoring and controlling of technical processes by a computer system.
A \textbf{\ac{HMI}} resides on level one of the \texttt{Industrial Automation Pyramid}
and allows an operator to interact with a machine or plant.
A \textbf{Historian} is a computer on level two, where conditions, input and output information is stored for long term analysis.
A \textbf{\ac{PLC}} is a device that is used to control a machine or plant and that is digitally programmed.
\textbf{Remote \ac{IO}} devices read inputs and write outputs over a field"=bus or network connection.

\subsection{Testbed Requirements}
As ~\cite{green2017pains} concluded, a testbed, sophisticated and versatile as it may be,
has little use unless it is broadly accessible.
Additionally, many scenarios should be able to be covered with LICSTER, resulting in the following requirements:

\begin{itemize}
    \item In order to make the testbed affordable for teaching and research, it must be designed for \textbf{low"=cost}.
    \item A \textbf{physical process} must be represented to study consequences of cyber attacks on \acp{ICS}.
    \item In order to obtain repeatable results, the entire process must be \textbf{reproducible}.
    \item The testbed must be \textbf{portable}, e.g. for teaching and demonstration to gain awareness.
          Furthermore, a small footprint is easier to handle, e.g. when modifying components.
    \item By using \textbf{open"=source} software and hardware, research on components is feasible.
    \item The testbed implementation should cover \textbf{level 0 to level 2} of a common \ac{ICS},
          focusing on the physical process and \ac{SCADA} environment.
\end{itemize}

\subsection{Related Work}\label{subsec:testbedrelatedwork}
Due to the ongoing interests in \ac{ICS} security a lot of testbeds were created around the world in the past years.
\cite{holm2015survey} described and compared 30 testbeds in a survey.
However, the testbeds listed there are either, expensive, closed"=source or virtualized.
LICSTER clearly distances itself from those testbeds with its clear focus on open"=source and low"=cost components.

Testbeds for \ac{ICS} are expensive, especially when they are built with standard hardware, for example the testbed used in
\cite{niedermaier18woot} 
is about 35 000 Euro.
This testbed fulfill the task of specific robustness tests of \ac{ICS} components.
However, the size and cost makes it unattractive for most researchers and students.

\cite{queiroz2009building} describe a modular testbed based on Modbus/TCP.
However, they only show simple \ac{DoS} attacks and make changes difficult because of proprietary hardware.

\cite{green2017pains} describe ten lessons learned by setting"=up an \ac{ICS} testbed.
They built up a huge testbed, but also concluded that local access and e.g. a mobile demo unit are essential.

\cite{mclaughlin2016cybersecurity} summarize the \ac{ICS} landscape and also describe the requirement of testbeds.
They highlight the need for real"=world physical consequences and their monitoring.
LICSTER matches these requirements and additionally enables monitoring the physical process.

\cite{maynard2018open} and also \cite{formby2018lowering} introduced an open framework for \ac{SCADA} virtualization and simulation.
However, a pure simulation or virtualization does not fulfill our requirements.
Nevertheless, this could be taken into consideration as an expansion.

\cite{foley2018science} use a Fischertechnik simulation model for cyber security science hackathons.
The basic idea is similar, but they are proprietary components and the testbed is significantly more expensive.

\subsection{Attacker Modeling}\label{subsec:attackermodel}
There is no single defense mechanism to mitigate all threats to a digital system.
Depending on the nature and origin of an attacker, some defenses might be less useful than others.
Therefore, before defining the possible attack scenarios against LICSTER, the potential attackers must be defined.

The \textbf{remote attacker} has network access to the \ac{ICS} through a router.
That means that the attacker can reach the system only via its \ac{IP} address and thus preventing attacks below the OSI
network layer three (\cite{day1983osi}).
This replicates the scenario of exposed control systems~(\cite{durumeric2015search}),
as for example, when devices are connected to the Internet for e.g. maintenance reasons.

In contrast to the capabilities of the remote attacker, the \textbf{local attacker} has direct access to the \ac{ICS}.
Being present at the plant site allows on the one hand the possibility for physical attacks on the individual \ac{ICS} components.
On the other hand, the direct access to the network switch where \ac{ICS} components are connected to.
This enables an attacker to perform \ac{ARP} spoofing and all the attacks that rely on it.

\subsection{Attack Scenarios} \label{subsec:lowcostattackscenarios}
A central distinction between traditional office networks and production networks is,
that \ac{ICS} networks has to manage the transition to physical processes.
That makes it all the more important for a relevant \ac{ICS} testbed to incorporate a physical process,
as it allows attacks on the system from an entirely different perspective.  
With LICSTER various attack scenarios on the \ac{ICS} levels zero to two are possible. 

The act of network \textbf{sniffing} can be separated into two methods.
The first is a passive approach.
An attacker can utilize a mirror port or network tap to capture the traffic or simply receive and read broadcast
messages.
The second method of network sniffing is an active technique, where traffic is redirected over the host of the attacker
by manipulation on the \ac{MAC} layer through e.g. \ac{ARP} poisoning.

More uncomplicated is the \textbf{\ac{DoS}} attack, where the target is flooded by network packages it needs to
react to. Is the amount of requests high enough, the accumulated network and/or \ac{CPU} load of reacting to each and
every package eventually causes the regular execution of the target to slow down or stop completely.

With a \textbf{\ac{MitM}} attack, an intruder manages to place himself between two communication partners through
manipulation of routing information on the IP layer or MAC layer. There the attacker is able to capture and/or
manipulate the exchanged packages.

Additionally, a \textbf{manipulation} over the network of \ac{ICS} components is possible,
as e.g. shown by \cite{niedermaier2017propfuzz} where the network interface of a \ac{PLC}
is fuzzed to manipulated the \ac{PLC}.

Apart from network based attack vectors, a culprit with \textbf{physical access} has a wide range of attacks at his
disposal such as manipulating devices, e.g. sensors, plug- and unplugging systems and straightforward destroying
components of the \ac{ICS}.
Following comes a certainly not comprehensive list of possible attacks mapped to the \ac{ICS} levels:

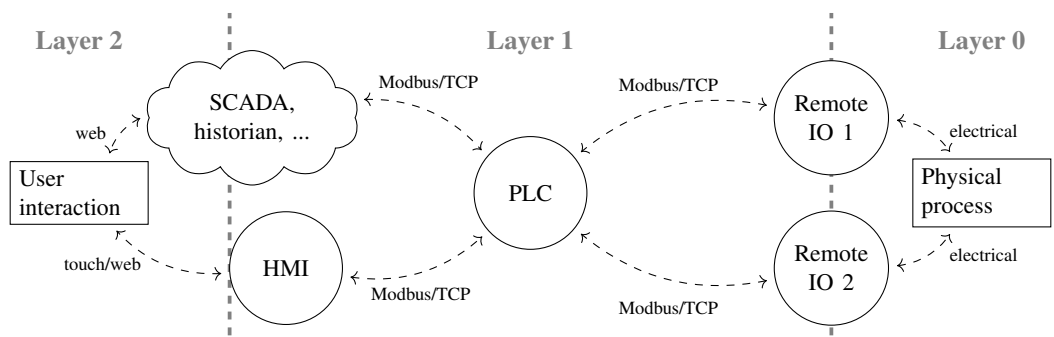
\begin{figure*}[htb]
  \centering
\begin{tikzpicture}[node distance=1cm, 
    shorten >= 3pt,shorten <= 3pt,
    auto]    
    
  \draw [-, dashed, line width=0.5mm,color=gray] (1,0) -- (1,4.5);
  \draw [-, dashed, line width=0.5mm,color=gray] (9,0) -- (9,4.5);
  
  \node [] at (-1,4) (x) {\textbf{\textcolor{gray}{Layer 2}}};
  \node [] at (5,4) (x) {\textbf{\textcolor{gray}{Layer 1}}};
  \node [] at (11,4) (x) {\textbf{\textcolor{gray}{Layer 0}}};
    
  \node [circle, draw, fill=white, text width=1.4cm, minimum size=1.0cm,
    inner sep=0.0cm, text centered] at (9,1) (n1) {\small Remote IO 2};
    
  \node [circle, draw, fill=white, text width=1.4cm, minimum size=1.0cm,
    inner sep=0.0cm, text centered] at (9,3) (n2) {\small Remote IO 1};
    
  \node [circle, draw, fill=white, text width=1.0cm, minimum size=1.5cm,
    inner ysep=0.1cm, text centered] at (5,2) (plc) {\small PLC};
    
  \node [circle, draw, fill=white, text width=1.0cm, minimum size=1.5cm,
    inner ysep=0.1cm, text centered] at (1.75,1) (hmi) {\small HMI};
    
  \node [cloud, draw, fill=white,cloud puffs=10,cloud puff arc=120, aspect=2, 
    text width=1.7cm,
    inner ysep=0.1cm, text centered] at (1.3,3) (c1) {\small SCADA, historian, ...};
  
  \node [draw, text width=1.6cm] at (11,2) (process) {\small Physical \\ \small process};
  \node [draw, text width=1.6cm] at (-1,2) (user) {\small User \\ \small interaction};
    
  \draw [<->, bend angle=25, bend left,dashed] (n1) 
    to node[below, text width=1.5cm, text centered,inner sep=0.3cm]{\scriptsize
		{Modbus/TCP}} (plc);
  \draw [<->, bend angle=25, bend right,dashed] (n2) 
    to node[above, text width=1.5cm, text centered,inner sep=0.2cm]{\scriptsize
		{Modbus/TCP}} (plc);          
  \draw [<->, bend angle=25, bend left,dashed] (c1) 
      to node[above, text width=1.5cm, text centered,inner
	  sep=0.3cm]{\scriptsize {Modbus/TCP}} (plc);
  \draw [<->, bend angle=25, bend right,dashed] (hmi) 
      to node[below, text width=1.5cm, text centered,inner
	  sep=0.2cm]{\scriptsize {Modbus/TCP}} (plc);
      
  \draw [<->, bend angle=25, bend left,dashed] (hmi) 
      to node[left, text width=1.5cm, align=right,inner sep=0.3cm]{\scriptsize {touch/web}} (user);
  \draw [<->, bend angle=25, bend right,dashed] (c1) 
      to node[left, text width=1.5cm, align=right,inner sep=0.2cm]{\scriptsize web} (user);
      
  \draw [<->, bend angle=25, bend right,dashed] (n1) 
      to node[right, text width=1.5cm, align=left,inner sep=0.3cm]{\scriptsize electrical} (process);
  \draw [<->, bend angle=25, bend left,dashed] (n2) 
      to node[right, text width=1.5cm, align=left,inner sep=0.3cm]{\scriptsize electrical} (process);
     \end{tikzpicture}
\caption{System view of LICSTER.}
\label{fig:lowcostnetworksetup}
\end{figure*}

Attacks on level 0 (process level):
\begin{itemize}[beginpenalty=10000] 
    \item Manipulating the physical process for example by removing the commodity
    \item Physically damage machines so that the process is not performed properly, e.g. with raw violence
\end{itemize}

Attacks on level 1 (field level):
\begin{itemize}[beginpenalty=10000]
    \item \ac{DoS} attack on e.g. sensors or actuators
    \item Interfering with availability by disconnecting the network or power plug
    \item Manipulating a sensor physically so it transmits spoofed values
    \item \ac{MitM} to manipulate the values between the \ac{PLC} and remote \ac{IO}
\end{itemize}

Attacks on level 1 (control level):
\begin{itemize}[beginpenalty=10000] 
    \item \ac{DoS} attack on the \ac{PLC} or \ac{HMI}
    \item Sniffing network traffic, e.g. to get sensitive production information
    \item \ac{MitM} to manipulate the values of the \ac{PLC} or \ac{HMI}
    \item Physical access to the \ac{HMI}, e.g. to stop the process
\end{itemize}

Attacks on level 2 (process control level):
\begin{itemize}[beginpenalty=10000] 
    \item \ac{DoS} attack on the \ac{SCADA} and historian systems
    \item Sniffing network traffic, e.g. to get sensitive process details
    \item \ac{MitM} to manipulating the values of the \ac{SCADA} or historian server
    \item Manipulating the state of the \ac{SCADA} system, e.g. to reduce the daily order count
\end{itemize}

Attacks in the various \ac{ICS} levels require different access rights and tools,
which are described in detail in \Cref{sec:evaluation} of the evaluation of LICSTER.

\section{Testbed Implementation}
\label{sec:implementation}
The testbed presented in this paper handles the physical, field, control and supervisory level of the \texttt{Industrial
Automation Pyramid} as described in \Cref{fig:lowcostnetworksetup}.
The implementation of the testbed is designed in a way to allow for industry~4.0 scenarios as well as the more traditional
\ac{ICS} cases. That means that the entire communication between the \ac{PLC}, the \ac{HMI} and even the remote~\acp{IO} is
based on TCP/IP protocols, namely Modbus/TCP.
The sensors themselves communicate with the remote~\acp{IO} on a fieldbus protocol, for which Modbus/TCP was chosen which is
broadly used in the industry.
An overview of the devices in the testbed is given in \Cref{tab:lowcost-devices} and described in detail in the
upcoming sections (\Cref{subsec:hmi} to \Cref{subsec:fischertechnik}).
The total amount of 577.-- Euro is not necessarily the cheapest choice, because it depends on the prices of the distributor.

\begin{table*}[htb]
    \centering
    \caption{Overview of devices used in our testbed. Prices are current prices on amazon.de} 
    \label{tab:lowcost-devices}
        {
        \begin{tabular}{l l l >{\ttfamily}l >{\ttfamily}r}
            \hline
            \textbf{Component} & \textbf{Software}     & \normalfont \textbf{Hardware} & \normalfont \textbf{IP}    & \normalfont \textbf{Price ca.}\\
            Remote IO          & FreeRTOS, LwIP        & Custom, STM32F7               & 192.168.0.51/52            & 79.-- Euro \\
            PLC                & OpenPLCv3             & Raspberry Pi 3                & 192.168.0.30               & 56.-- Euro \\
            SCADA              & ScadaLTS, Logging     & Raspberry Pi                  & 192.168.0.20               & 56.-- Euro \\
            HMI                & Custom, PyModbus      & Raspberry Pi with Display     & 192.168.0.10               & 139.-- Euro \\
            Switch             & -                     & TP-Link                       & 192.168.0.1                & 32.-- Euro \\
            Process            & -                     & Fischertechnik                & -                          & 195.-- Euro \\
            Others             & -                     & e.g. cables                   & -                          & 20.-- Euro \\
           \hline
           Total               &                       &                               &                            & 577.-- Euro \\
        \end{tabular}%
        }
\end{table*}

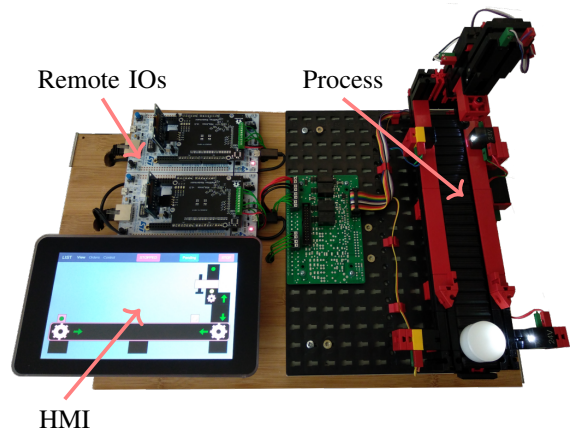
\begin{figure}[htb]
\centering
    \begin{tikzpicture}
	\node[inner sep=0pt] (setup) at (0,0)
            {\includegraphics[width=0.45\columnwidth]{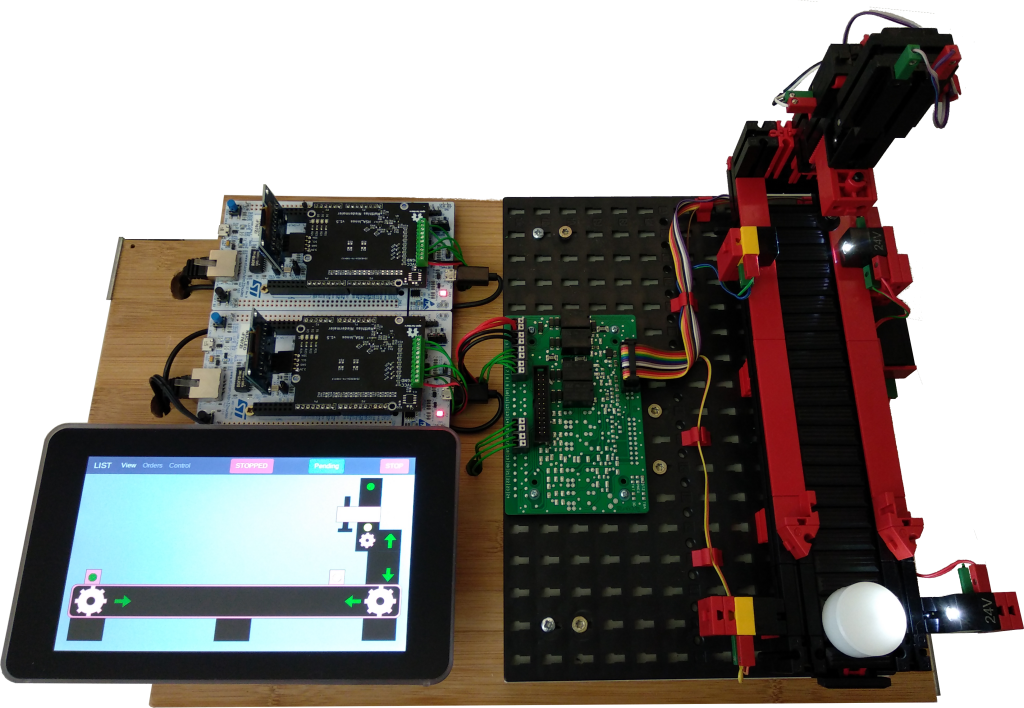}};
    
    \draw [<-, color=red!50, line width=0.5mm] (-2,-1.5) 
      to (-3,-2.7) node[below, text width=1.5cm, align=center, color=black]{HMI};
    \draw [<-, color=red!50, line width=0.5mm] (-2,0.5) 
      to (-2.5,1.3) node[above, text width=1.9cm, align=center, color=black]{Remote IOs};
   \draw [<-, color=red!50, line width=0.5mm] (2.3,0.0) 
      to (0.7,1.3) node[above, text width=1.8cm, align=center, color=black]{Process};
    \end{tikzpicture}
	\caption{Top view on LICSTER.}
	\label{fig:lowcostsetup}
\end{figure}

\Cref{fig:lowcostsetup} shows the testbed mounted on a wooden board for portability.
The \ac{HMI}, the two remote~\acp{IO} and the physical process are placed on top.
The network switch and two Raspberry~Pis are fixed on the bottom side of the board.

\subsection{\ac{PLC}} \label{subsec:plc}
As \ac{PLC} the open"=source solution OpenPLC from \cite{alves2014openplc} is used.
It is a soft \ac{PLC}, which means that it can be run on various operating systems and on hardware with and also without \acp{IO}.
Within LICSTER OpenPLC runs on a Raspberry~Pi, with our \ac{PLC} procedure programmed in
\ac{ST} which is uploaded to the \ac{PLC} over its own web portal.

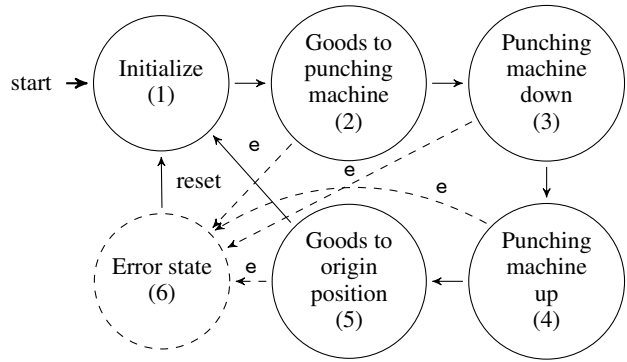
\begin{figure}[H]
    \centering
    \footnotesize
    \begin{tikzpicture}[>=stealth',shorten >=2pt,auto,shorten <=2pt,node distance=0.55 cm,
        every state/.style={align=center,minimum size=1.8cm},
        every edge/.style={draw,thick},
        loop label/.style={draw,align=center,text width=2cm,outer sep=4pt,minimum height=1cm}
        ]

        \node[initial,state] (A)                {Initialize \\ (1)};
        \node[state]         (B) [right= of A]  {Goods to \\ punching \\ machine \\ (2)};
        \node[state]         (C) [right= of B]  {Punching \\ machine \\ down \\ (3)};
        \node[state]         (D) [below= of C]  {Punching \\ machine \\ up \\ (4)};
        \node[state]         (E) [left= of D]  {Goods to \\ origin \\ position \\ (5)};
        \node[state, dashed] (F) [left= of E]  {Error state \\ (6)};

        \node[rotate=0] at (0.5,-1.3) {reset};

        \draw[->] ([xshift=0ex]A.east) -- ([xshift=0ex]B.west);
        \draw[->] ([xshift=0ex]B.east) -- ([xshift=0ex]C.west);
        \draw[->] ([xshift=0ex]C.south) -- ([xshift=0ex]D.north);
        \draw[->] ([xshift=0ex]D.west) -- ([xshift=0ex]E.east);
        \draw[->] ([xshift=0ex]E.north west) -- ([xshift=0ex]A.south east);
        \draw[->] ([xshift=0ex]F.north) -- ([xshift=0ex]A.south);
        
        \draw[->, dashed] ([xshift=0ex]B) -- node[above,text width=0.1cm]{\texttt{e}\\~} ([xshift=0ex]F);
        \draw[->, dashed] ([xshift=0ex]C) -- node[above]{\texttt{e}} ([xshift=0ex]F);
        \draw[->, dashed, bend right] (D.north west) to node[right]{~~~~~~~~~~~~~\texttt{e}} (F.north east);
        \draw[->, dashed] (E) -- node[above]{\texttt{e}} ([xshift=0ex]F);
    \end{tikzpicture}
    \caption{Program sequence of the process implemented on the \ac{PLC}.}
    \label{fig_lowcoststates}
\end{figure}

The program contains a simple, repeatable process which can be triggered
and monitored by the \ac{HMI} as shown in \Cref{fig_lowcoststates}.
The process consists of five stages in which it moves and processes a plastic cylinder as an workpiece example.

\circledp{1}~Before starting, the initial conditions must be fulfilled with everything at rest and the cylinder at its place.
\circledp{2}~The conveyor belt moves the plastic cylinder to the punching machine and stops right underneath it.
\circledp{3}~The punching machine moves downwards until its lower limit switch is triggered,
\circledp{4}~then continues to move up again until its upper limit switch is triggered.
\circledp{5}~Finally, the conveyor belt moves the plastic cylinder away from the punching machine, back to its origin and then stops.
\circledp{6}~If an detectable error (\texttt{e}) occurs or the emergency stop is pressed,
the machine goes in an error state, by stopping every movement.
This can be reset on the \ac{HMI}.

\subsection{\acf{HMI}} \label{subsec:hmi}
\begin{figure}[htb]
\centering
    \begin{tikzpicture}
	\node[inner sep=0pt] (hmi) at (0,0)
            {\includegraphics[width=0.55\columnwidth]{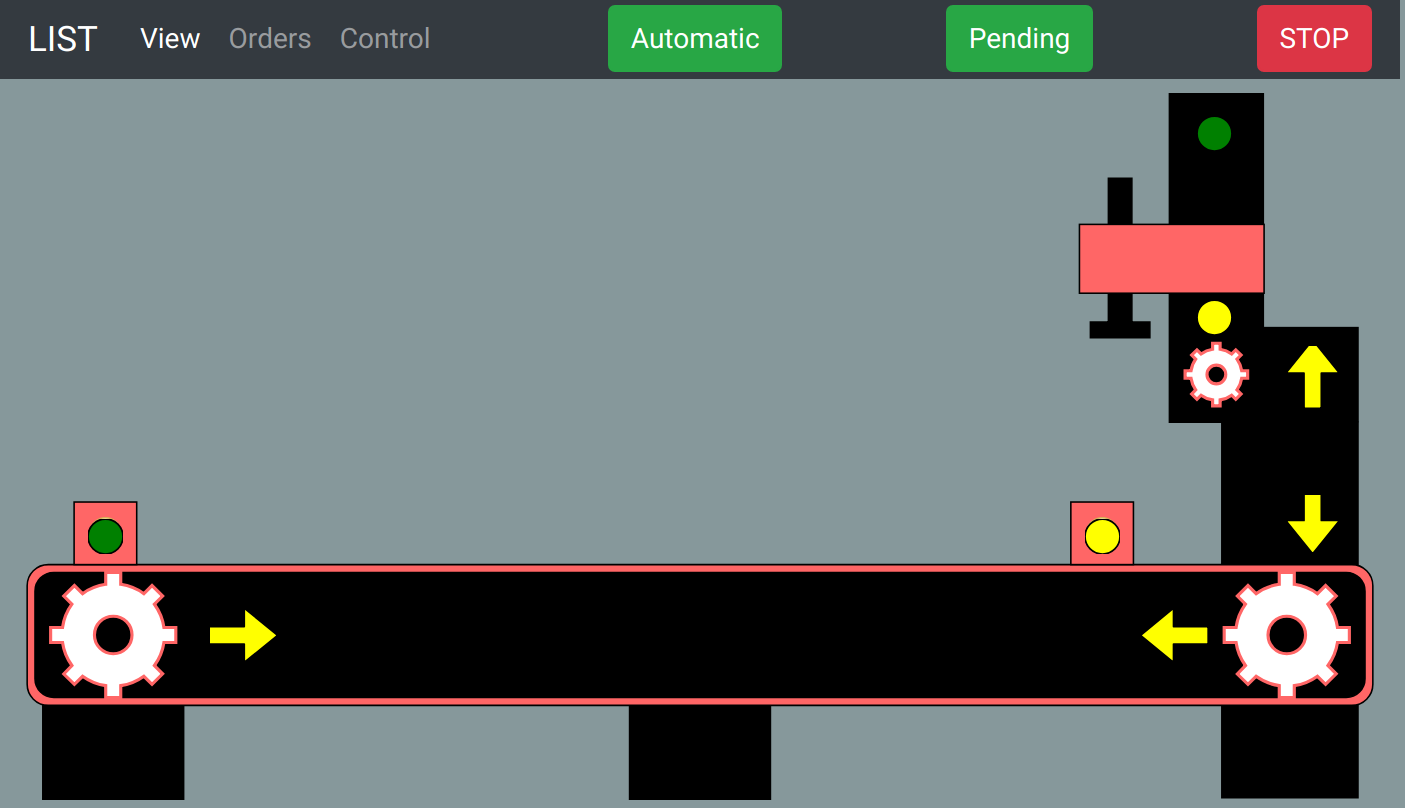}};
    \end{tikzpicture}
	\caption{Picture showing the \ac{HMI}.}
	\label{fig:lowcosthmi}
\end{figure}

The \ac{HMI} is provided by a webserver which runs on a dedicated Raspberry~Pi attached to a touch"=screen.
The web application is split into three areas -- view, order and control -- where the user has the possibility to monitor the
process, place orders and thus trigger the described process, and to manually control the conveyor belt and the punching
machine via the touch panel.
This behaviour resembles a real \ac{HMI}.
\Cref{fig:lowcosthmi} shows a screenshot of the \ac{HMI}, where the process is observed. 
Through this functionality and usage, authentic attack scenarios can be set up.
To base the \ac{HMI} on a webserver has several perks: Firstly, it reflects the change of technology introduced by industry~4.0.
Companies like Siemens are already pursuing this approach with their WinCC/Web
Navigator\footnote{\url{https://w3.siemens.com/mcms/human-machine-interface/en/visualization-software/scada/wincc-options/wincc-web-navigator/Pages/Default.aspx}}.
Secondly, knowledge about web"=technologies is wide"=spread and conveniently accessible, making this \ac{HMI} easy to
understand, extend and exploit.
And thirdly, with small modifications, such as introducing Wi"=Fi to the Raspberry Pi, the
\ac{HMI} can be effortlessly ported to tablets or smartphones, which introduces new attack vectors and, again, represents the
shift to contemporary technologies.

Additionally to monitoring the process, orders can be placed and the controller can be set in a manual mode, where the
process is controlled by the operator.
This also brings the human component into play in the testbed.

\subsection{Remote \acs{IO}} \label{subsec:remoteio}
The remote \acp{IO} are self"=developed, open"=source solutions based on a development board from STMicroelectronics.
The base board is a STM32F767ZI\footnote{\url{https://www.st.com/en/evaluation-tools/nucleo-f767zi.html}}
with an Arduino Uno V3 header, where our custom add"=on board is connected to.
Using the Arduino Uno V3 header for the custom \ac{PCB} allows the underlying prototyping board to be changed easily
with other compatible boards.
This is relevant, for example, if higher performance, an energy"=saving solution or cheaper hardware is needed.

\begin{figure}[htb]
\centering
    \begin{tikzpicture}
	\node[inner sep=0pt] (display) at (0,0)
            {\includegraphics[width=0.47\columnwidth]{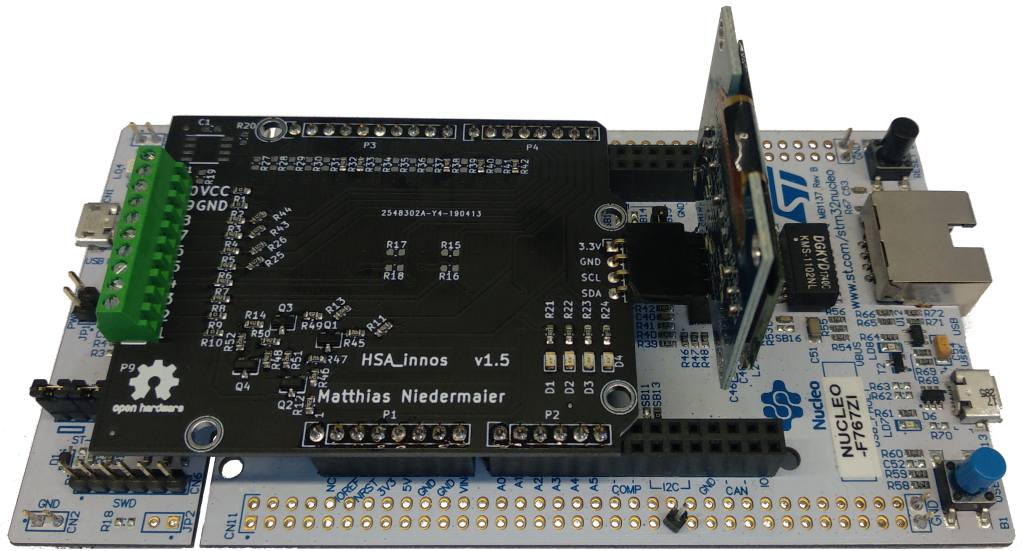}};   
    \draw [<-, color=red!50, line width=0.5mm] (-1.8,-0.7) 
      to (-2.0,-2.1) node[line width=0.05mm, below, align=center, color=black]{\small Transistors};   
    
    \draw [<-, color=red!50, line width=0.5mm] (0.5,-0.8) 
      to (1.0,-2.1) node[line width=0.05mm, below, align=center, color=black]{\small LEDs};  
    
    \draw [<-, color=red!50, line width=0.5mm] (3.0,0.4) 
      to (2.4,-2.1) node[line width=0.05mm, below, align=center, color=black]{\small Network};  
      
    \draw [<-, color=red!50, line width=0.5mm] (1.7,0.6) 
      to (2.4,2.1) node[line width=0.05mm, above, align=center, color=black]{\small Display}; 
      
    \draw [<-, color=red!50, line width=0.5mm] (-2.9,0.3) 
      to (-2.4,2.1) node[line width=0.05mm, above, align=center, color=black]{\small IOs}; 
      
    \draw [<-, color=red!50, line width=0.5mm] (-1.0,0.3) 
      to (-1.0,2.1) node[line width=0.05mm, above, align=center, color=black]{\small Custom PCB}; 
      
    \draw [<-, color=red!50, line width=0.5mm] (1.0,1.0) 
      to (0.3,2.5) node[line width=0.05mm, above, align=center, color=black]{\small STM32 development board}; 
    \draw [<-, color=red!50, line width=0.5mm] (-3.2,0.5) 
      to (-2.8,2.5) node[line width=0.05mm, above, align=center, color=black]{\small USB}; 
    \end{tikzpicture}
	\caption{\ac{PCB} of the remote \acp{IO}.}
	\label{fig:lowcostpcb}
\end{figure}
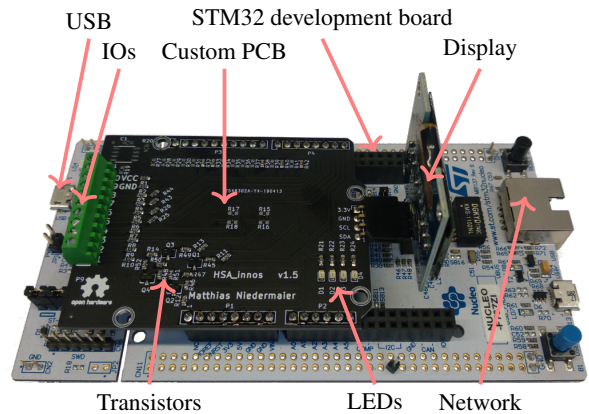

\begin{figure}[htb]
\centering
    \begin{tikzpicture}
	\node[inner sep=0pt] (display) at (0,0)
            {\includegraphics[width=0.25\columnwidth]{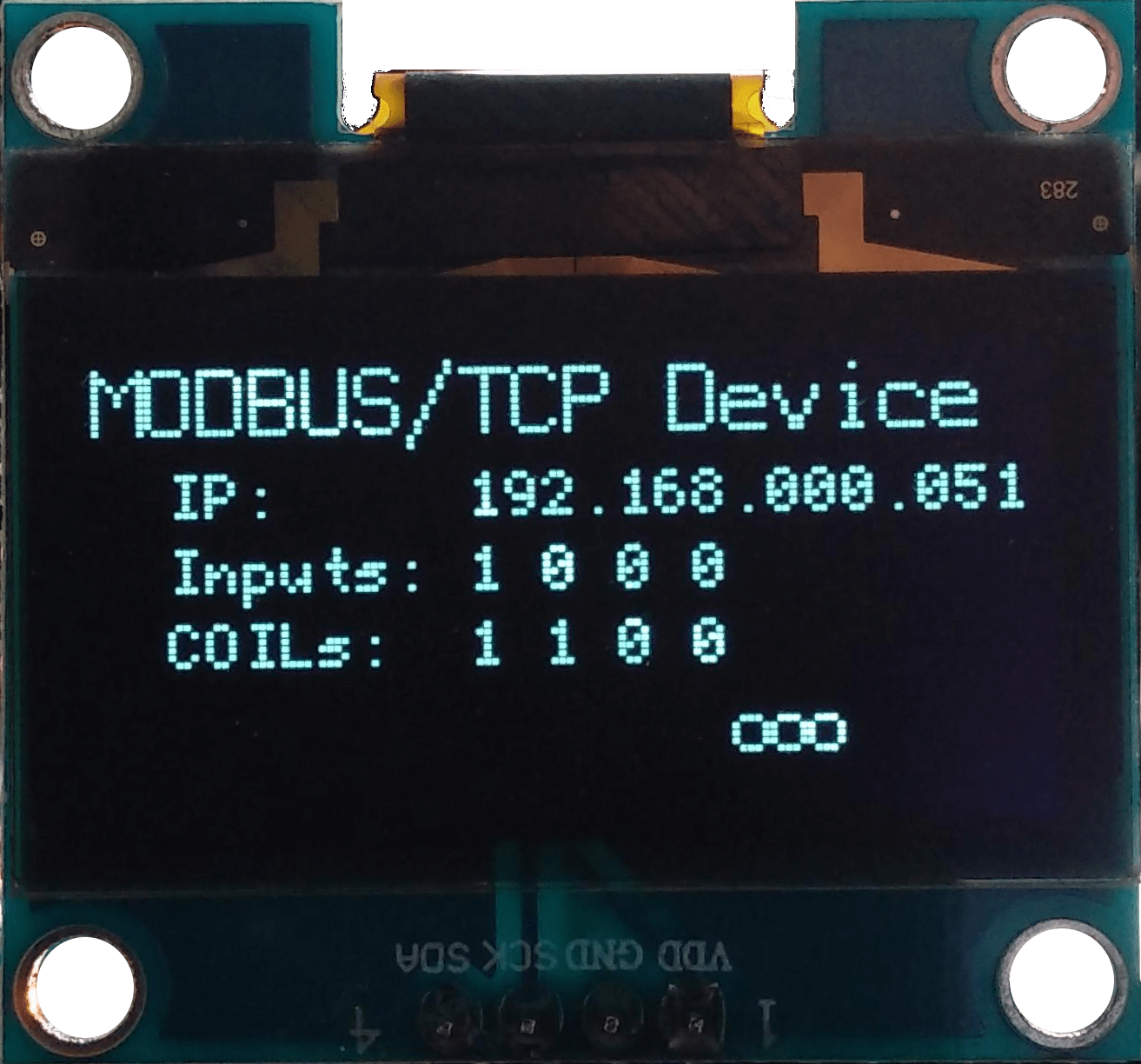}};
    \end{tikzpicture}
	\caption{Picture showing the display mounted on each remote \ac{IO}.}
	\label{fig:lowcostdisplay}
\end{figure}

The \ac{USB} connector for programming and power is placed on the left side of the development board.
The RJ45~Ethernet connector for networking is mounted on the right side.
The custom \ac{PCB} is mainly necessary to convert the 24V of the physical process 
to the 3.3V of the STM32 development board.
Additionally, each remote~\ac{IO} is connected to a display as illustrated in \Cref{fig:lowcostdisplay}.
There the operator can monitor the current state of the Modbus/TCP inputs registers and coils.

Applying displays to field level components is a trend 
that can also be seen by real"=world components. 
In larger plants and systems, it simplifies identifying erroneous devices for the operator.
What is more, these devices often allow a simple on"=site basic configuration during commissioning.

\subsection{Physical Process (Fischertechnik)} \label{subsec:fischertechnik}
\begin{figure}[htb]
\centering
    \begin{tikzpicture}
	\node[inner sep=0pt] (fischertechnik) at (0,0)
            {\includegraphics[width=0.47\columnwidth]{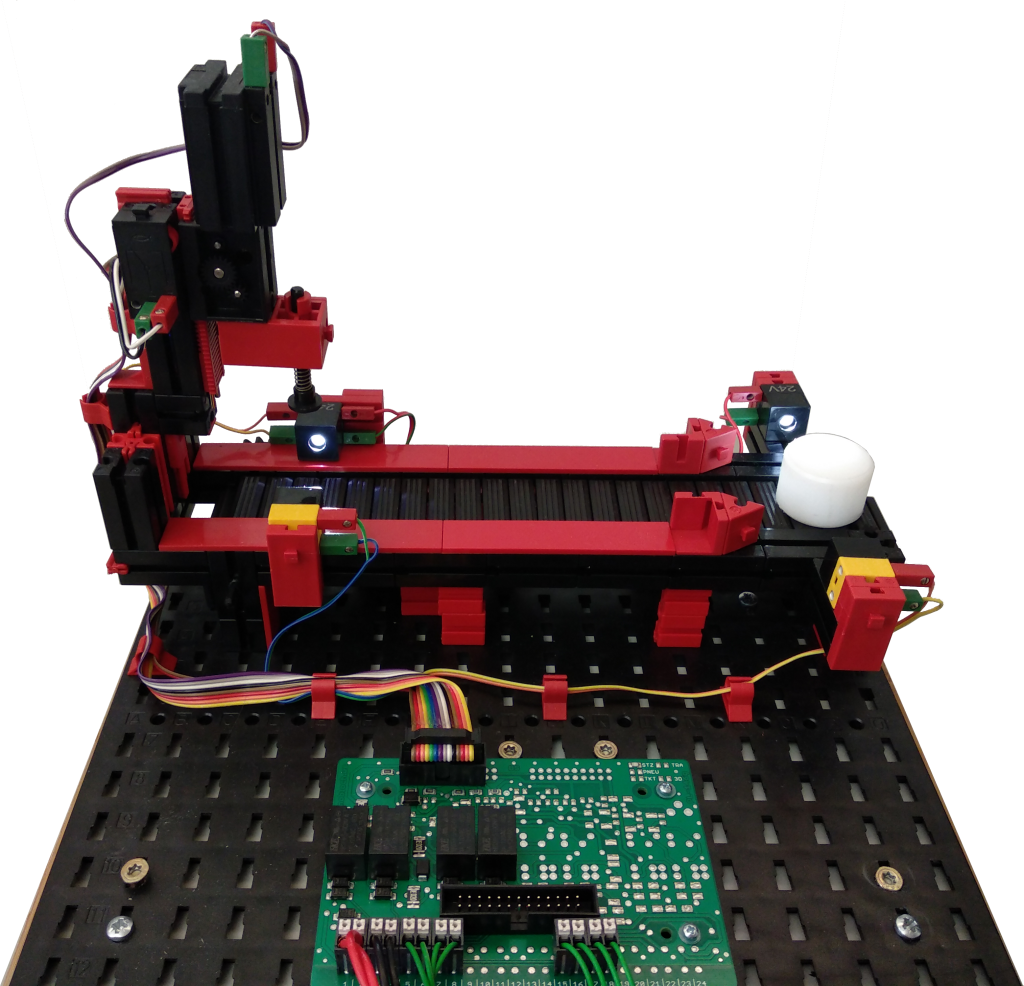}};
    
    \draw [<-, color=red!50, line width=0.5mm] (2.3,0.0) 
      to (0.6,0.6) node[line width=0.05mm, above, align=center, color=black]{\small Goods};
      
    \draw [<-, color=red!50, line width=0.5mm] (2.1,0.6) 
      to (1.4,1.4) node[line width=0.05mm, anchor=east, text width=2.0cm, align=right, color=black]{\small Light barrier};
    \draw [<-, color=red!50, line width=0.5mm] (-1.5,0.5) 
      to (-0.6,1.4);
    
    \draw [<-, color=red!50, line width=0.5mm] (0.0,-2.5) 
      to (-0.5,-3.9) node[line width=0.05mm, anchor=north, align=center, color=black]{\small Driver board};
      
    \draw [<-, color=red!50, line width=0.5mm] (0.5,-0.1) 
      to (2.5,-3.9) node[line width=0.05mm, anchor=north, align=center, color=black]{\small Conveyor belt};
      
    \draw [<-, color=red!50, line width=0.5mm] (-2.3,2.1) 
      to (-0.9,2.9) node[line width=0.05mm, anchor=west, align=center, color=black]{\small Punching machine};
    
    \draw [<-, color=red!50, line width=0.5mm] (-2.7,1.7) 
      to (-0.9,2.25) node[line width=0.05mm, anchor=west, text width=2.0cm, align=left, color=black]{\small Limit switch};
    \draw [<-, color=red!50, line width=0.5mm] (-2.7,0.7) 
      to (-0.9,2.15);
    \end{tikzpicture}
	\caption{Picture showing the Fischertechnik setup.}
	\label{fig:lowcostfischertechnik}
\end{figure}
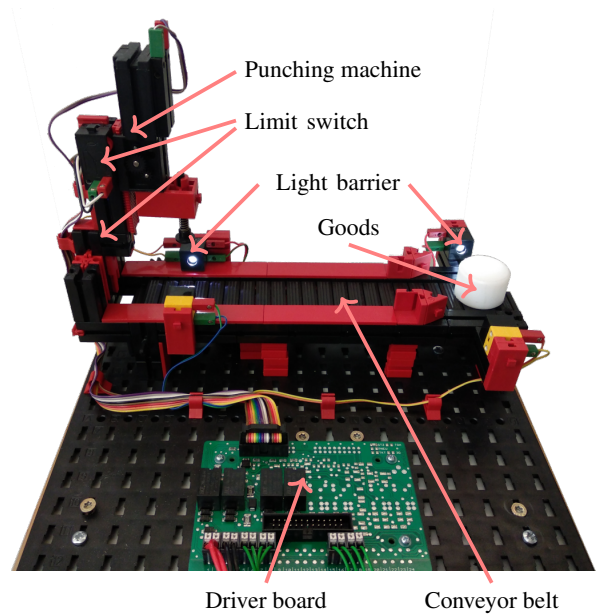

One of the main requirements is to use a real physical process so that impacts are directly visible.
However, in order to keep the necessary skills low, an affordable as well as manageable solution is used in our testbed.
Another requirement to the process is, that it should be automatically repeatable without the need for human
interaction.
That way the researcher or student has enough time to execute his attacks and to observe the effects.
The selected device is a Fischertechnik punching machine
96785\_sim\footnote{\url{https://www.fischertechnik.de/en/products/simulating/training-models/96785-sim-punching-machine-with-conveyor-belt-24v-simulation}},
as shown in \Cref{fig:lowcostfischertechnik} which costs 195 Euro.

The Fischertechnik system consists out of a conveyor belt, two light"=barrier, two limit"=switches and two motors. The
light"=barriers are placed at each end of the conveyor belt and the limit"=switches control the upper and lower limit of
the punching machine. One electric motor drives the conveyor belt clockwise and counterclockwise and the other electric
motor lifts and lowers the punching machine from and to the conveyor belt.

\begin{figure*}[htb]
\centering
    \begin{tikzpicture}
	\node[inner sep=0pt] (display) at (0,0)
            {\includegraphics[width=0.95\textwidth]{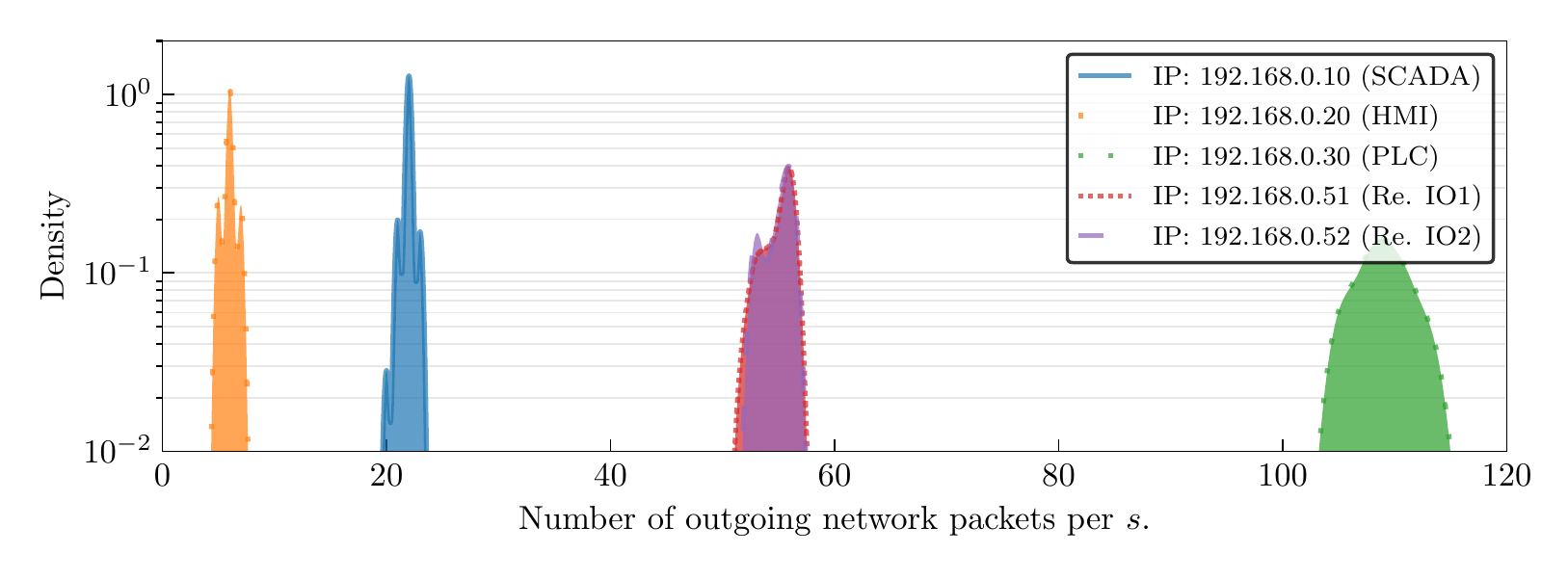}};            
    \end{tikzpicture}
    \vspace*{-0.5cm}
	\caption{Density plot showing the number of packets per second of each device within LICSTER.}
	\label{fig:lowcostnetworkdensity}
\end{figure*}

\subsection{SCADA/Historian} \label{subsec:scada}
The software used as \ac{SCADA} system for our testbed is Scada"=LTS\footnote{\url{https://github.com/SCADA-LTS/Scada-LTS}}.
It is an open"=source solution which supports Modbus/TCP and is entirely web"=based.
It also represents the shift to contemporary technologies in industry~4.0.
The software runs on a Raspberry~Pi.
Its webpage can be accessed by any system within the network.

\subsection{Necessary Skills} \label{subsec:skills}

Although the testbed is designed to keep the entry hurdle for new students as low as possible, basic knowledge about the
following three aspects is necessary.
\circled{1} Basic electric knowledge is required to safely connect a few wires to the system.
However, this is very limited, since only four sensors and two motors must be connected.
\circled{2} When the STM32"=based remote \acp{IO} should be used,
soldering skills and a soldering iron are necessary.
This can be avoided by e.g. using Raspberry Pis with 24V \acp{IO} as recommended by the OpenPLC project,
but this results in a smaller testbed and also a more expensive price.
\circled{3} Basic Linux skills are of benefit to set"=up the Raspberry Pis
and use the attacking tools.
However, we provide necessary material and guides to setup the testbed,
which reduce the initial hurdles to a minimum.

\section{Evaluation and Benchmarking of the Testbed}
\label{sec:evaluation}
The evaluation of our testbed consists of three tiers.
At first it is assured that the overall concept is viable and the communication between the components works as expected.
Secondly, the previously identified attacks are systematically applied to the testbed and evaluated for their
feasibility and effects to the system.
Finally, it is evaluated what open research questions could be assessed with the help of the proposed testbed.  

\subsection{Evaluation of the Implementation} \label{subsec:lowcostconcept}
For some research questions and learning content, network traffic is a critical element. 
For example, intrusion detection can be configured and evaluated based on the network traffic captures of executed attacks.
\Cref{fig:lowcostnetworkdensity} shows the density distribution of outgoing packets per second per
component during a one minute capture while the testbed was running its process.

It clearly shows that the amount of packets per second differs, depending on the device.
The \ac{PLC}~(192.168.0.30), which contains most of the logic and therefore is the most communicative component with
about 110 packets per second.
This is because the \ac{PLC} polls the remote~\acp{IO} every 100~$ms$ while simultaneously being polled by the \ac{SCADA}
and \ac{HMI}.
In comparison, the communication of the \ac{HMI}, which updates the values only once every 500~$ms$, amounts to significantly less traffic.
Either remote \acp{IO} takes about 55 packets per second to communicate, which is mostly due to the constant polling of the \ac{PLC}.
The similar behaviour of the remote~\acp{IO} leads to a correlative density representation.
Lastly, the \ac{SCADA} system, which has no hard timing requirements, only amounts to about 20 packets per second. 

\subsection{Attack Validation within the Testbed}
\begin{table*}[htb]
    \centering
    \caption{Evaluation of a selection of possible devices in the testbed.} 
    \resizebox{.99\textwidth}{!}{%
    \label{tab:lowcost-attacks}
        \begin{tabular}{c | l c c c c l l l l}
            \hline
            \cell{\textbf{ICS} \\ \textbf{Level}}    & \cell{\textbf{Attack} \\ \textbf{description}} & \cell{\textbf{CIA}} & \cell{\textbf{STRIDE}} & \cell{\textbf{Remote} \\ \textbf{attacker}} & \cell{\textbf{Local} \\ \textbf{attacker}} & \cell{\textbf{Tools}} & \cell{\textbf{Skill} \\ \textbf{level}}  & \cell{\textbf{Impact}} &  \cell{\textbf{Detection} \\ \textbf{difficulty}} \\
            \circledg{1}  &  \hspace{8mm}\circledg{2} & \circledg{3} & \circledg{4} & \multicolumn{2}{c}{\circledg{5}} & \multicolumn{1}{c}{\circledg{6}} & \multicolumn{1}{c}{\circledg{7}} & \multicolumn{1}{c}{\circledg{8}} & \multicolumn{1}{c}{\circledg{9}}\\
            \hline 
            \multirow{3}{*}{\cell{0 \\ process \\ level}}& Manipulate     & A &
			TD & \xmark          & \cmark          & ---          & low  & high
			& easy \\ & Physically Damage    & A & TD & \xmark          & \cmark          & ---          & low  & high & easy \\ \\
            \hline
            \multirow{4}{*}{\cell{1 \\ field \\ level}}& \acs{DoS} Sensor & A & D & \cmark          & \cmark          & hping3        & low  & high & easy \\ & Disconnect IO power/network & A & TD & \xmark          & \cmark          & ---          & low   & high & easy \\ & Manipulate IO physical & A & TD & \xmark          & \cmark          & ---          & low  & high & easy \\ & \ac{MitM} spoof values IO-PLC  & CIA & STRIDE & \xmark          & \cmark          & script          & high  & high & medium \\
            \hline
            \multirow{5}{*}{\cell{1 \\ control \\ level}}& \ac{DoS} PLC & A & D & \cmark          & \cmark          & hping3        & low  & high & easy \\ & \ac{DoS} HMI  & A & D & \cmark          & \cmark          & hping          & low  & medium & easy \\ & Sniffing network & C & I & \xmark          & \cmark          & Tcpdump          & low  & low & difficult \\ & \ac{MitM} spoof values HMI-PLC  & CIA & STRIDE & \xmark          & \cmark          & script          & high  & high & medium \\ & Physical access \ac{HMI}  & CIA & STRIDE & \xmark          & \cmark          & ---          & low  & low & medium \\
            \hline
            \multirow{4}{*}{\cell{2 \\ process \\ control \\ level}}& \ac{DoS} SCADA   & A & D &
            \cmark          & \cmark          & hping3        & high  & low & easy \\ & Sniffing
            network & C & I &\xmark           & \cmark          & Tcpdump          & low  & low
            & difficult \\ & \ac{MitM} spoof values SCADA-PLC & CIA & STRIDE & \xmark          &
            \cmark          & script          & high  & high & medium \\ & Attack \ac{SCADA} & CIA &
            STRIDE & \cmark         & \cmark          & script          & medium   & high & medium \\
        \end{tabular}%
    }
\end{table*}
\Cref{tab:lowcost-attacks} shows an overview of selected attacks performed on the testbed.
In order to correlate the attacks with the \textbf{\ac{ICS} levels}~\circledg{1}, the first column enumerates the levels
zero to two of the \texttt{Industrial Automation Pyramid}.
The second column \textbf{Attack description}~\circledg{2} of the table contains a list of attacks derived from
\Cref{subsec:lowcostattackscenarios}.
These are the scenarios for a potential attacker.
The three basic protection goals of IT"=Security are \acl{CIA}.
Therefore, the third column, \textbf{\acs{CIA}}~\circledg{3}, maps each attack to the protection goals it compromises.
The \textbf{STRIDE}~\circledg{4} threat model by \cite{kohnfelder1999threats} in the fourth column uses the indicators
\textbf{S}poofing, \textbf{T}ampering, \textbf{R}epudiation, \textbf{I}nformation disclosure, \textbf{D}enial of service
and \textbf{E}levation of privilege as a more detailed threat mapping.
Next, the column five maps the \textbf{attacker model}~\circledg{5} introduced in \Cref{subsec:attackermodel}
to each attack, to clarify which attacks actually need physical access and which do not.
Column 6 contains \textbf{Tools}~\circledg{6} which are used to evaluate and perform attacks on the testbed are.
Where viable ready"=made and open"=source solutions are preferred. Additionally, customized scripts are provided to
execute attacks where no ready to use tools is available.
Here is a brief overview over the tools and the corresponding attacks:
\begin{itemize}
    \item Network scanning is done with \texttt{nmap} by \cite{lyon2009nmap}.
        The specific Modbus/TCP \ac{NSE} script\footnote{\url{https://nmap.org/nsedoc/scripts/modbus-discover.html}} 
          is used to identify Modbus/TCP devices.
      \item For network sniffing the Linux tool \texttt{tcpdump}~(\cite{jacobson1989tcpdump}) and
          \texttt{wireshark}~(\cite{combs2008wireshark}) is used.
          Wireshark, a software for traffic capture and protocol dissection for various protocols including Modbus/TCP,
          allows for easy package analysis.
    \item For flooding and \ac{DoS} attacks the tool \texttt{hping3}~(\cite{sanfilippo2005hping3}) is used.
          It supports a multitude of protocols and configurations to perform different forms of \ac{DoS} attacks.
    \item For \ac{MitM} attacks a custom python script for easy editing is used, based on the libraries pymodbus and
        scapy~(\cite{biondi2011scapy}).
    \item For active manipulation of values in the Modbus/TCP communication, a custom python script or interactive
        python shell with pymodbus is employed.
\end{itemize}

The seventh column, \textbf{Skill level}~\circledg{7}, refers to the knowledge required by the attacker to achieve his
malicious goals.
Each attack scenario is rated from low, medium and high which primarily represents the amount of system specific insight an
attacker needs to follow through with his attack.
Other aspects, such as the basic technical knowledge and the expertise in available tools do not weigh in the rating
quite as much since most tools and relevant documentation are freely and sufficiently available.
To measure the \textbf{Impact}~\circledg{8} in column eight, again a rating from low, over medium to high is used to
reflect the consequences to the system and the physical process of each attack.
The severity of the rating depends on factors such as whether or not the plant can be damaged as a direct or indirect
result of the attack, e.g. when the punching machine does not stop at the limit switch and crashes into the ground.
Finally, the \textbf{Detection difficulty}~\circledg{9} is assessed in column nine in the three levels low, medium high,
as mentioned previously.
It represents a rough estimation of the likelihood of successful detection of an attack by defensive mechanisms, e.g. an \ac{IDS}.
The skill level, impact and detection difficulty are rated in three levels.
Attacks in real scenarios depend on many circumstances and can vary heavily when compared with each other.
This is why the table can only give a tendency, which should be taken with a grain of salt.

With this evaluation, we have shown that even low"=cost testbeds offer many possibilities of attacks.
It is important to see direct effects for new researchers and students, such as the physical process.
This shows \ac{ICS} specifics, as digital devices interacts with the real world and attacks on network devices can have an impact on a process.

\subsection{Discussion of Extensions and Research Questions} \label{subsec:researchquestions}
This section elaborates on how LICSTER's functionality can be enhanced by extensions and what research questions could be
assessed on the foundation of our testbed.
Also the research ideas introduced by \cite{cardenas2008research} are picked up in this paper as well.
This demonstrates how adaptive LICSTER really is and how much room for research and discussion it provides, despite its
simplicity.
In fact, it is this very simplicity that makes this testbed so easy to use and promising for students and beginners.

\subsubsection{Extensions}
LICSTER can easily be extended, for example by virtual \ac{ICS} components~(\cite{antonioli2015minicps}) or standard
office clients in virtual machines. The extensibility of our testbed allows for the simulation of potentially huge
environments, always with the physical process integrated.
Apart from enhancing LICSTER by further components and clients, its communication capabilities can be extended by
additional protocols such as, for example, OPC UA.
That way LICSTER can be used as platform to evaluate the security aspects of upcoming \ac{ICS} protocols (\cite{renjie2010research}).
Evaluations like these could lead to the establishment of requirements for secure protocols in \acp{ICS}.
One topic that is being discussed, e.g. by \cite{givehchi2014control}
is to operate parts of an \ac{ICS}, like \acp{PLC}, in the cloud or fog.
It is interesting to examine further security research in addition to the evaluation of availability and control timings.
All what is necessary is to move the OpenPLC, which runs on any Linux"=based computer, into the cloud or fog.

\subsubsection{Offense Scenarios}
\cite{morris2013industrial} presented 17 attacks against \acp{ICS} which use the same Modbus/TCP protocol as LICSTER.
Hence, these attacks can be executed on our testbed and evaluated with the real world impacts on the physical process.
One of the most serious dangers to critical infrastructures is an \ac{APT}, as explained, among others,  by~\cite{gouglidis2018protecting}.
LICSTER provides an elementary but sufficient platform to further investigate these types of attacks, since it provides
all of the relevant components of an \ac{ICS} including the physical process. 

\subsubsection{Defense Mechanisms}
A protection mechanism often used for \acp{ICS} is network monitoring (\cite{zhu2010scada}).
Due to the fact that LICSTER spans over multiple \ac{ICS} layers, it incorporates various types of network
communication, as can be seen in \Cref{subsec:lowcostconcept}.
For example, the communication between the \ac{PLC} and the remote \acp{IO} shows clear timing criticality
while the traffic between the \ac{SCADA} system and the \ac{PLC} does not.
With these distinctive communication characteristics, LICSTER can be used to evaluate \ac{IDS} implementations and to
test their detection mechanisms.  
Also the remote \acs{IO}, can be programmed with custom firmware, which is mostly not possible when proprietary hardware is used.
With this, intelligent \ac{IIoT} edge nodes, e.g. for intrusion detection, can be placed into the testbed.\\
This small selection of selected topics shows that it is often not the size that counts,
but it is more important to map an entire industrial process within a testbed.
This allows a simple demonstration of impacts in the physical world caused by cyber attacks.
Particularly for learning purposes, it is important to have simple tools to assess complex topics.

\section{Conclusion}
\label{sec:conclusion}

In this paper, we presented LICSTER, an open"=source, low"=cost \ac{ICS} testbed for education and research.
We have shown, that the concept can be and set"=up for about 500~Euro,
This way we lower the entry barrier so that more people can get hands"=on experience with \acp{ICS}.
To enhance the learning experience of how the physical world interacts with the digital environment, we introduced
suitable attacker models and possible attacks.
With these measures we intent to provide easy access to more students and researchers to the topic of \ac{ICS} security.

Furthermore, even with a testbed as elementary as LICSTER, current and relevant research questions can be assessed, due to
the open"=source nature of the project and its components.
With LICSTER offensive as well as defensive techniques can be tested and evaluated on different \ac{ICS} levels.
The physical process is a key segment of LICSTER, which allows for a haptic understanding of the effects of cyber attacks on \acp{ICS}.

\section*{Availability} \label{sec:availability}
We provide supplementary material at \url{https://github.com/hsainnos}.
The material includes all required source code, how"=tos, binaries, scripts and network captures to understand and reproduce the results.

\bibliography{\jobname}

\end{document}